# NEWS ON DARK MATTER IN GALAXIES AND CLUSTERS

Françoise Combes
DEMIRM, Observatoire de Paris
61 Av. de l'Observatoire, F-75 014 Paris, France

Daniel Pfenniger
Observatoire de Genève, CH-1290 Sauverny, Switzerland

**Abstract.** Major progresses have been made this last year towards a better knowledge of the invisible mass. Michel Spiro will talk in details about the micro-lensing experiments and their promising results; the ROSAT satellite has provided extended X-ray maps of the hot gas, which traces dark matter in galaxy clusters: they reveal lower amounts of dark matter in clusters than was previously derived; the dark to visible mass in clusters is not larger than its value in spiral galaxies. It was shown, by X-ray data and gravitational lenses analysis that the dark matter density is highly peaked towards the cluster centers. A new dark matter candidate has also been proposed, in the form of cold and fractal molecular gas that could be present around most late-type spiral galaxies and account for the observed flat rotation curves.

# I – ROTATION CURVES OF SPIRAL AND IRREGULAR GALAXIES

The best evidence until now of the existence of dark matter is provided by the HI gas rotation curves of spiral galaxies. Already, the optical rotation curves obtained with ionised gas (H$\alpha$) were found to be surprisingly flat (e.g. Rubin et al. 1980), but they do not extend enough outside the optical disk to unambiguously require the presence of dark matter. On the contrary, the radio rotation curves obtained up to 3–4 optical radii with HI at 21cm, are maintained flat so far out that the presence of a dark component is unavoidable (e.g. Bosma 1981, van Albada et al. 1985).

## A – HUBBLE SEQUENCE

More recently, it has been realised that the fraction of dark matter depends on the morphological type: it is increasing along the Hubble sequence, from Sa to Sd (also called "early" and "late" galaxies respectively), and maximises for dwarf irregulars. The Hubble sequence was until now characterised by five main parameters monotonously varying from Sa to Sd: 1) the decreasing bulge-to-disk luminosity ratio, indicating that the mass concentration decreases along the sequence; 2) the increasing gas fraction; 3) the decreasing total mass; 4) the decreasing fraction of heavy elements, processed in stars; and 5) the increasing pitch angle, i.e., the spiral arms being more wound for early-type galaxies.

From enhanced sensitivity HI observations in gas-poor early-type galaxies, falling rotation curves have been discovered (Casertano & van Gorkom 1991, Salucci & Frenk 1989). This means that early-type galaxies *have proportionally less dark matter.*

On the other side of the sequence, dwarf irregulars are observed with rising rotation curves (e.g. Carignan & Freeman 1988). Dwarfs spirals *have proportionally more dark matter.* Since the Hubble sequence is also a mass (or luminosity) sequence, it emerges that the dark-to-luminous mass decreases with total visible mass (Persic & Salucci 1993).

These new results should be re-enforced with larger statistics. It is of prime importance, for instance, to find several galaxies with no evidence at all of dark matter. This will provide strong constraints on non-baryonic dark matter models, and even rule out modified Newtonian gravity hypotheses.

The Hubble sequence is not only a classification of galaxies according to their morphology; the galaxy disks may evolve in a short time-scale (less than $10^9$ yrs) along the sequence, through numerous dynamical instabilities, such as the formation and dissolution of spiral waves and bars (Pfenniger et al. 1994). The non-axisymmetric waves produce gravity torques that drive the mass towards the center: a galaxy then evolves from late to early types (Sd to Sa), increasing its luminous mass, and its mass concentration; at the same time, star-formation consumes the gas and increases the heavy element content. The evolution cannot be the other way around, since mass concentration, star formation, and nucleosynthesis are irreversible processes. The fact that during the evolution, the luminous mass increases, while the dark mass fraction decreases suggests that the dark matter has been transformed into stars, particularly because disk galaxies like the Galaxy are too fragile structures to afford a substantial mass accretion in or at the periphery of the optical disk (Tóth & Ostriker 1992). This is only possible if the dark matter around galaxies is composed mainly of hydrogen and helium gas, not yet condensed in compact objects.

## B – MAXIMUM DISK

How well do we know the radial distribution and exact amount of dark matter in spiral galaxies? How much mass is associated with visible stars?

The $M/L$ mass-to-light ratio for the stellar population depends on age, metallicity, and the overall colours. Tinsley (1981) has computed the variations of $M/L$ with (B-V) colour, from 1 to 10 for red to blue galaxies. She deduced already at that time that since the total (dynamical) mass to luminosity ratio was not varying as much, the fraction of dark matter was higher in blue galaxies, i.e. in the late-types of the Hubble sequence. The possibility to vary $M/L$ for stars, according to their age and colour, makes the maximum disk hypothesis tenable for most galaxies: i.e., it is possible to explain the rotation curve in the inner parts of the galaxy by maximising the contribution of the visible matter in fitting $M/L$. No dark matter is then required within the optical disk. Recently, Buchhorn et al. (1994) have succeeded to fit the observed H$\alpha$ rotation curves of 97% galaxies in a sample of about 500: they compute the gravitational potential of the galaxy from the red image (less perturbed by extinction), assuming a constant $M/L$ as a function of radius. The derived rotation curve fits surprisingly well the kinematical H$\alpha$ data; it remains quite flat, except for small-scale wiggles, tracing the streaming motions associated with spiral arms. Of course, the choice of a lower $M/L$ for the stars is still possible, within the uncertainties; however the precise fits of these wiggles is a strong argument in favor of the maximum disk, since any spherically distributed dark matter will not follow so tightly the spiral arm streaming motions.

When the rotation curve is flat, the radial distribution of dark matter corresponds to an isothermal component, the density varying as $r^{-2}$ in the extended parts of the disk, and the projected surface density as $r^{-1}$. In some cases, however, not only isothermal distributions, but also $r^{-3}$ or $r^{-4}$ asymptotic laws for the density are possible (Lake & Feinswog 1989).

The main argument against the maximum disk hypothesis rises within non-baryonic dark matter scenarios. It is indeed difficult to imagine a hollow dark matter component, with a central hole filled in by the baryonic matter, if the dark and luminous matter components are physically different and independent (one is spherical, the other one flattened). However, this argument is no longer valid if both types of matter are two forms belonging to the same physical component, and simply one form converts into the other one within some radius.

### C – CONSPIRACY

The flatness of most rotation curves has been considered a puzzle, since it implies a conspiracy between two supposedly independent dynamical components: the visible disk and the spherical dark halo. We have seen already that this flatness is no longer universal, rotation curves having tendencies to rise for late-types and fall for early-type galaxies. But yet, most intermediate types have a flat curve.

When the curve is flat already within the optical disk, the visible matter alone can account for it (e.g. Freeman 1993). The problem is to explain the continuity of the curve in the outer parts, dominated by dark matter. Several scenarios have been proposed, arguing that dark matter is dragged by baryonic infall (Blumenthal et al. 1986). But in dark matter dominated dwarfs the predicted core radius of the total mass is too short with respect to the observations (Moore 1994).

The conspiracy is of course no longer a problem if the dark and luminous matter are two aspects of the same dynamical component, one phase being progressively transformed in the other, during galaxy evolution.

## II – CLUES FROM THE GAS

### A – RADIAL DISTRIBUTION OF GAS AND DARK MATTER

The atomic gas serves as a tracer of dark matter, since only HI is visible in the outer parts of spiral disks. There is moreover some evidence that the gas and dark matter are intimately related. From the flat rotation curves, the surface density of dark matter $\sigma_{DM}$ varies asymptotically as $1/r$, and the HI surface density $\sigma_{HI}$ is also observed to vary in $1/r$. Bosma (1981) was the first to notice a constant ratio of $\sigma_{DM}/\sigma_{HI}$ as a function of radius in spiral galaxies. The constant ratio has been confirmed by many authors (Sancisi & van Albada 1987, Puche et al. 1990), and varies between 10 and 20 according to morphological type (Broeils 1992).

## B – 3D DISTRIBUTION: POLAR RINGS AND THICKNESS OF HI PLANES

Albeit many observational efforts, it has not been possible until now to determine with certainty the 3D-shape of the dark matter distribution.

Polar rings, i.e., material orbiting almost perpendicular to the main plane of the galaxy were a good hope to test the 3D distribution of mass, and in particular the shape of dark matter. Much work has been devoted to this problem but with controversial results (Whitmore et al 1987; Sackett & Sparke 1990). The modelling is in fact confronted with many difficulties, the main one being intrinsic to the formation mechanism: gas settling in the polar ring and then forming stars. Due to gas clouds collisions, a polar ring cannot form in a galaxy that has already a gaseous component at the same radii in the equatorial plane. Therefore, the equatorial velocities are not determined at the same radius as the polar ones, and the flattening of the halo cannot be directly determined (Reshetnikov & Combes 1994).

Warps and their problematic longevity have long been an argument in favor of spherical dark halos, since the time-scale of differential precession is then considerably enlarged. However, the warp lines of nodes are not observed to be wound as they should even at radii where the disk dominates the mass, and other explanations involving gas accretion and short life-time are now considered more plausible (Binney 1992).

As well as the HI kinematics in the galaxy plane being used to test the radial distribution of mass, the HI velocity dispersion and height can be used to test the mass distribution perpendicular to the plane. In the isothermal sheet model of a thin plane, the height $h(r)$ of the gaseous plane is $h(r) = \sigma_v^2(r)/\pi G \mu(r)$, where $\sigma_v(r)$ is the $z$-velocity dispersion, and $\mu(r)$ the total surface density, within the gaseous plane. Since the HI gas $z$-velocity dispersion has been observed constant in the outer parts of face-on galaxies (Shostak & van der Kruit 1984; Dickey et al. 1990), the HI characteristic height as a function of radius gives the amount of dark matter $\mu(r)$ included within the HI plane. In the Milky Way, the height $h(r)$ is increasing linearly with radius (Merrifield 1992), and its precise value is that expected if all the dark matter was flattened at least as much as the flaring HI plane. In M31, Brinks & Burton (1984) also discovered a flaring linear with radius.

## C – GAS RESERVOIRS AROUND MOST SPIRAL GALAXIES

It has long been known that intermediate-type spiral galaxies have formed stars at about a constant rate along the Hubble time (Kennicutt 1983). Infall of gas has been invoked to solve the problem of gas consumption by star formation (Larson et al. 1980), and even the maintenance of spiral structure needs the replenishment of the dissipative component (Toomre 1990). Recently Braine & Combes (1993) have found that interacting galaxies appear to possess more visible gas than isolated ones, as if galaxy interactions would suddenly reveals more gas. This could be due to the driving of the gas towards the center by tidal torques, if there exists an outer gas reservoir.

# III – A NEW CANDIDATE FOR DARK MATTER: COLD H$_2$

All these arguments about galaxy evolution and gas properties led us to propose a new model for dark matter around spiral galaxies (Pfenniger, Combes, & Martinet 1994; Pfenniger & Combes 1994). The idea of diffuse gas has been eliminated since the beginning by the Gunn-Peterson test (Gunn & Peterson 1965): if the Universe was filled by an homogeneous diffuse gas, we should detect it in the spectrum of remote quasars, as an absorption trough starting at the redshift of the quasar. This test could be done for atomic gas (21cm line), or ionised gas (Ly$\alpha$) etc...

But the gas could be in a clumpy and fragmented structure as it is observed in nearby molecular clouds. It has been proposed by several authors that the interstellar medium is distributed in a fractal structure, over at least 4 orders of magnitude in scale (e.g., Scalo 1985; Falgarone et al. 1992). We extrapolate this fractal down by 2 orders of magnitude in scale, since structures as small as $10-20\,\mathrm{AU}$ are detected in the interstellar gas (HI VLBI interferometry: Diamond et al. 1989, proper motions when passing in front of quasars: Fiedler et al. 1987, or pulsars: Cognard et al. 1993). The physical structure of this fractal will be detailed in another contribution in this conference. By physical arguments, we deduce that the building blocks of the fractal, the "clumpuscules", have masses of the order of Jupiters ($10^{-3}\,M_\odot$), volumic densities of $10^{10}\,\mathrm{H\,cm^{-3}}$, and surface densities of $10^{25}\,\mathrm{H\,cm^{-2}}$. The gas is in molecular form, and in thermal equilibrium with the 3 K cosmic background.

## IS THAT GAS OBSERVABLE AROUND GALAXIES?

Most of the mass is in the form of H$_2$ molecules, which unfortunately do not radiate at low excitation, since by symmetry they have no dipole moment (no emission of rotational lines). In molecular clouds, H$_2$ is traced by the CO emission. This is highly dependent on metallicity, and CO emission is expected to disappear far away from the optical disk, where the gas is deficient in heavy elements. High-energy $\gamma$ rays are also a good tracer of all nucleons, since they are produced in the interaction between cosmic rays and matter. However, the main sources of cosmic rays are supernovae and other stars, and therefore this tracer is again restricted to the optical disk.

Since the gas is cold, any emission would not be easily detected against the background, and may be the best hope is to detect it in absorption. Already many absorption lines corresponding to the HI atomic gas are detected in the spectrum of quasars: along some line of sights, up to 100 Ly$\alpha$ absorbing systems have been detected, and represent the well-known Ly$\alpha$ forest. These absorption lines correspond to moderate column densities, between $10^{15}$ to $10^{20}\,\mathrm{H\,cm^{-2}}$; if they are associated to galaxies they imply an equivalent cross-section, corresponding to a radius of about $R = 200 - 400\,\mathrm{kpc}$, at $z = 2.5$ (Sargent 1988). UV observations with the Hubble Space Telescope have recently revealed that the absorbing systems are still numerous at low redshift (Morris et al. 1991; Bruhweiler et al. 1993). These absorbing systems could be the envelopes of dense and massive molecular clumps.

At the present temperature of the background radiation (2.7 K), it is highly probable that some of the H$_2$ has become solid. At the average density of the clumpuscules, the phase transition occurs at 3 K. If H$_2$ ice is present in the shape of snow flakes, they can be efficiently coupled to the background radiation, and radiate themselves like a black-body. This will also occur in the presence of a small amount of dust. The emission of these particles could then be detected by faint fluctuations of the microwave background.

Let us mention also that these clumpuscules are not compact enough to produce micro-lensing for stars in the Large Magellanic Clouds. However, on the line of sight of a remote quasar (at least at 100 Mpc distance), intervening clumpuscules or larger clumps of the fractal structure could produce, by micro-lensing, the observed non-intrinsic variability of these quasi-punctual sources (cf. Hawkins 1993).

# IV – GALAXY CLUSTERS

The X-ray emission of the hot gas in galaxy clusters has been explored with enhanced sensitivity by the satellite ROSAT, for 4 years now. More clusters have been detected, and in the brightest clusters, the hot gas has been studied at a much larger distance from the cluster center. Within the hypothesis of hydrodynamical equilibrium, the hot gas is a tracer of the cluster potential, and the total mass has been determined at larger cluster radii. A comparative analysis of groups and clusters have shown that the total mass-to-light ratio decreases with the size of the object (David et al. 1994). For the largest clusters, which contain also the greatest fraction of hot gas, the $M/L$ ratio begins to decreases, the dark matter becomes luminous. This $M/L$ ratio never rises much above that for spiral galaxies. Previous estimates of larger $M/L$ relied on virial estimates, and on the hypothesis of isotropic motions in the outer parts of clusters, which is not verified since these outer parts are not relaxed.

The radial distribution of dark matter has been determined more accurately from a deeper analysis of X-ray data. The results are consistent with a very small (unresolved) core radius of dark matter: much smaller than that of the hot gas (Gerbal et al. 1992; Briel et al. 1992; Durret et al. 1994). This is confirmed independently by the gravitational lens method for total mass determination, when arcs and mini-arcs are present (Hammer 1991, Wu & Hammer 1993). This more concentrated distribution suggests that the dark matter is baryonic in order to dissipate energy and cool to a lower temperature than the hot gas (Briel et al. 1992).

## A – COOLING FLOWS

The hot gas in the center of clusters reaches the density threshold for thermal instability, and its cooling time falls below the Hubble time: it then cools down very quickly (Rees & Ostriker 1977). The rate of cooling has been measured, of the order of $\dot{M} = 10^2$ $M_\odot$/yr (e.g. Fabian 1987). But the cooled gas has never been detected, a puzzle that several scenarios have tried to get round, including star formation with biased initial mass functions. Recently, X-ray absorption has revealed the existence of a large mass of gas in the center of clusters (White et al. 1991; Allen et al. 1993).

Since this gas is not seen in HI (emission, absorption, O'Dea et al. 1994), it could be hidden in the form of molecular clouds (White et al. 1991; Ferland et al. 1994; Fabian, this meeting). Several tentatives of CO detection have been made in vain (Grabelsky & Ulmer 1990; Mc Namara & Jaffe 1994; Antonucci & Barvainis 1994; Braine & Dupraz 1994). The low temperature of the molecular gas, and its low filling factor, if it is distributed in small dense clumps, can easily explain these non-detections.

In the hypothesis of cold gas as dark matter, the hot gas comes from the heating of some of the cold gas of the outer parts of galaxies during cluster formation. Since the mass of this cold gas in late-type galaxies is one or two orders of magnitudes than the visible mass, it is easy to explain the domination of the hot gas mass in clusters, between 2 and 8 times the galaxy visible masses (Edge & Stewart 1991). During cooling flows the gas resumes its cold phase, and falls down the cluster potential well, accounting for the small core radii.

## B – BARYON FRACTION OF THE TOTAL MASS OF A CLUSTER

The new X-ray data allow to make a more accurate mass account within galaxy clusters. From stellar luminosity, hot X-ray gas, the visible baryonic matter is already a fraction $0.15\,h_{50}^{-3/2}$ of the total mass (White et al. 1993).

But standard big-bang nucleosynthesis constrain $\Omega_b$ below $0.09\,h_{50}^{-2}$ (Smith et al. 1993). This is hardly compatible with $\Omega = 1$, unless most of the non-baryonic dark matter has been maintained outside of clusters, by some formation mechanism. White et al. (1993) show that this cannot be accounted for by dissipative effects during cluster formation, and conclude:
- either that $\Omega$ is much lower than 1,
- or the theory of standard big-bang nucleosynthesis is wrong.

Let us mention that a model of universe, where dark matter is only baryonic, is not ruled out by observations. In X-ray clusters, the visible mass, including the hot gas, is only 10 times lower than the total dynamical mass. The total $\Omega$ inferred is of the order of 0.05, still compatible with the standard big-bang nucleosynthesis.

## V – CONCLUSIONS

The enhanced sensitivity of ROSAT X-ray data have not revealed the presence of more dark matter on large scales. In fact, the dark matter becomes visible in rich clusters as hot emitting gas. This implies that the fraction of baryonic matter in clusters becomes larger and larger, incompatible with the joint hypothesis of standard big-bang nucleosynthesis (BBN) and $\Omega = 1$. On the other hand, at the galaxy scale, the amount of baryons is still insufficient to meet the lower bound of BBN, so that there must exist some baryonic dark matter. This could exist in the shape of brown dwarfs, as might have been seen by micro-lensing experiments (e.g. Spiro, this meeting). Or this baryonic dark matter could be in the form of cold molecular gas in the outer parts of galaxies. This gas will then be available for star formation, all along the galaxy life-time, explaining galaxy evolution along the Hubble sequence. The fractal structure of this medium is simply the generalisation of what is observed in nearby molecular clouds at slightly higher temperatures: $5-15$ K. The cold gas is in statistical equilibrium with the cosmic background radiation at 2.7 K. This new candidate has the advantage of providing a natural origin of the hot gas in clusters, and explaining the observed high baryonic fraction in strong X-ray clusters.